\documentclass[conference]{IEEEtran}
\IEEEoverridecommandlockouts

\usepackage{subcaption}  
\usepackage{color}
\usepackage{pifont}
\usepackage{bbm}
\usepackage{stfloats}
\usepackage[short]{optidef}
\usepackage{multirow}
\usepackage{comment}

\usepackage{amsmath,amssymb,amsthm}
\usepackage{ragged2e} 

\usepackage{array}
\IEEEoverridecommandlockouts

\usepackage{cite}
\usepackage{amsmath,amssymb,amsfonts}
\usepackage{algorithmic}
\usepackage{graphicx}
\usepackage{textcomp}
\usepackage{xcolor}
\usepackage{amsmath}

\def\BibTeX{{\rm B\kern-.05em{\sc i\kern-.025em b}\kern-.08em
    T\kern-.1667em\lower.7ex\hbox{E}\kern-.125emX}}
\begin{document}
\title{Active RIS-Assisted URLLC NOMA-Based 5G Network with FBL under Jamming Attacks
}

 \author{\IEEEauthorblockN{Ghazal Asemian, Mohammadreza Amini, Burak Kantarci}\\  \vspace{-0.15in}
 \IEEEauthorblockA{School of Electrical Engineering and Computer Science, University of Ottawa, Ottawa, ON, Canada \\
 \texttt{\{gasem093,mamini6,burak.kantarci\}@uottawa.ca}}
 \vspace{-0.35in}
 }

\maketitle

\begin{abstract}
In this paper, we tackle the challenge of jamming attacks in Ultra-Reliable Low Latency Communication (URLLC) within Non-Orthogonal Multiple Access (NOMA)-based 5G networks under Finite Blocklength (FBL) conditions. We introduce an innovative approach that employs Reconfigurable Intelligent Surfaces (RIS) with active elements to enhance energy efficiency while ensuring reliability and meeting latency requirements. Our approach incorporates the traffic model, making it practical for real-world scenarios with dynamic traffic loads. We thoroughly analyze the impact of blocklength and packet arrival rate on network performance metrics and investigate the optimal amplitude value and number of RIS elements. Our results indicate that increasing the number of RIS elements from 4 to 400 can improve signal-to-jamming-plus-noise ratio (SJNR) by 13.64\%. Additionally, optimizing blocklength and packet arrival rate can achieve a 31.68\% improvement in energy efficiency and reduced latency. These findings underscore the importance of optimized settings for effective jamming mitigation.\end{abstract}


\section{Introduction} \label{Sec:Introduction} 
A jamming attack involves the intentional transmission of interference signals by a malicious entity to disrupt legitimate communication \cite{Lohan.2024}. Particularly in Non-Orthogonal Multiple Access (NOMA) systems. This can impair the Successive Interference Cancellation (SIC) decoding process and power allocation, degrading service quality for users with weaker signals \cite{dao2023dealing}. Uplink jamming, in particular, can significantly increase error rates \cite{zeng2020power}. One mitigation technique for uplink multiple-input multiple output (MIMO)-NOMA systems is to reduce total transmit power using precoders at transmitters and equalizers at the base station (BS) \cite{wang2019power}. However, this approach may fall short in environments with severe multipath fading or obstacles. To enhance communication reliability, the combination of power allocation and strategic access point (AP) positioning, such as deploying an Unmanned Aerial Vehicle (UAV) as AP, is explored in \cite{dao2021defeating}. The authors also discuss using Reinforcement Learning (RL) to adapt to dynamic conditions such as source node positions and jamming information in uplink NOMA scenarios under jamming attacks \cite{dao2023dealing}.
However, these studies rely on Shannon capacity, which assumes infinite blocklength—an impractical condition for real-world scenarios with finite transmission blocklengths \cite{zhong2024minimum}. Systems with strict latency requirements need short blocklength transmissions \cite{zhong2024minimum}. Optimizing power allocation and sub-channel assignments under finite blocklength (FBL) conditions is another jamming mitigation technique for NOMA systems \cite{amini2024bypassing}. The study in \cite{bello2023optimal} explores optimal power allocation under both asymptotic and FBL conditions for interference management in uplink NOMA, emphasizing the importance of FBL for low-latency applications. However, anti-jamming methods like power control, beamforming, and using powerful transmitters or active relays result in high power consumption \cite{yang2020intelligent}. Additionally, the assumption of access to jamming information and its channel state is unrealistic in practical scenarios.

A RIS consists of numerous elements that adjust the phase and amplitude of incoming signals, enhancing spectral efficiency, especially when a direct propagation path is unavailable \cite{yuan2021reconfigurable}. This improves signal quality without increasing transmit power \cite{yang2021energy}. There are three types of RIS elements based on their architecture: passive RIS, which only alters the phase; absorptive RIS, which adjusts both phase and amplitude to attenuate the signal; and active RIS, which also modifies phase and amplitude but amplifies the incoming signal.
The impact of passive RIS on downlink wireless communication under jamming and eavesdropping attacks is explored in \cite{sun2021intelligent}, where the authors address imperfect channel state information (CSI) and unknown jamming beamforming. Their method optimizes active beamforming at the base station and passive beamforming at the RIS to maximize system rate and minimize information leakage. In \cite{sun2022robust}, jamming mitigation is examined using a game-theoretic approach. The authors employ a Bayesian Stackelberg game to tackle imperfect angular CSI, resulting in a robust beamforming design.
The integration of NOMA into wireless communication enhances spectral and energy efficiency \cite{khan2019joint}. While RIS has been used to improve physical layer security (PLS) in NOMA networks, research on mitigating jamming attacks in RIS-based NOMA networks is limited. A study \cite{tabeshnezhad2023ris} explored using RIS in uplink NOMA under smart jamming attacks to improve quality-of-service (QoS) while controlling transmit power, focusing on two types of RIS elements: regular phase-shift elements, which reflect signals, and absorptive RIS, which suppress jamming signals.

In addition to absorptive RIS, active RIS is gaining interest in improving PLS due to its ability to provide effective robustness while enhancing the quality of the desired signal \cite{ji2023securing}.
Authors in \cite{sun2023active} explore a downlink MIMO network under a multi-jammer attack. The study proposed a novel receiver architecture involving an active-passive cascaded RIS to benefit from the capabilities of both types of elements. The passive RIS elements nullify the jammer while the active RIS elements amplify the desired signal.

As the uplink communication is more vulnerable and critical to jamming attack, this work focuses on RIS-based uplink NOMA communication. 
Our communication scheme includes an uplink NOMA network in FBL considering transmission diversity. Our main goal is to maximize the energy efficiency while meeting reliability and latency requirements for ultra-reliable low latency (URLLC) applications. The main contributions of this work are as follows.
\begin{itemize}
    \item Derive network metrics such as users' reliability, packet delay, and energy efficiency in an active RIS-assisted NOMA-based uplink communication scheme under jamming attack. We further consider finite blocklength and transmission diversity to reflect the URLLC. 
    \item Formulate an optimization model to obtain the user equipment (UE)'s optimum allocated transmit power alongside other network parameters such as blocklength, RIS element phases and amplification factors, and the number of packet re-transmissions to mitigate the jamming effect.
    
\end{itemize}
This work is further unique in that it takes into account both the data traffic behavior of users and the noise from active RIS elements in its derivations. Based on our findings, increasing the number of RIS elements results in significant growth in the value of the signal-to-jamming-plus-noise ratio (SJNR) which directly impacts the reliability. On the other hand, with an increased number of RIS elements, the required amplitude to achieve URLLC-required reliability decreases significantly which enables the system to achieve higher energy efficiency. Furthermore, the optimization of blocklength and packet arrival rate can reduce target latency and enhance energy efficiency. 

\vspace{-1mm}
\section{System Model} \label{Sec:System_model}
\subsection{Network Model}
\begin{figure}[htbp]
    \centering
    \includegraphics[trim={3.5cm 2cm 5.16cm 2.41cm},clip , width=.85\linewidth]{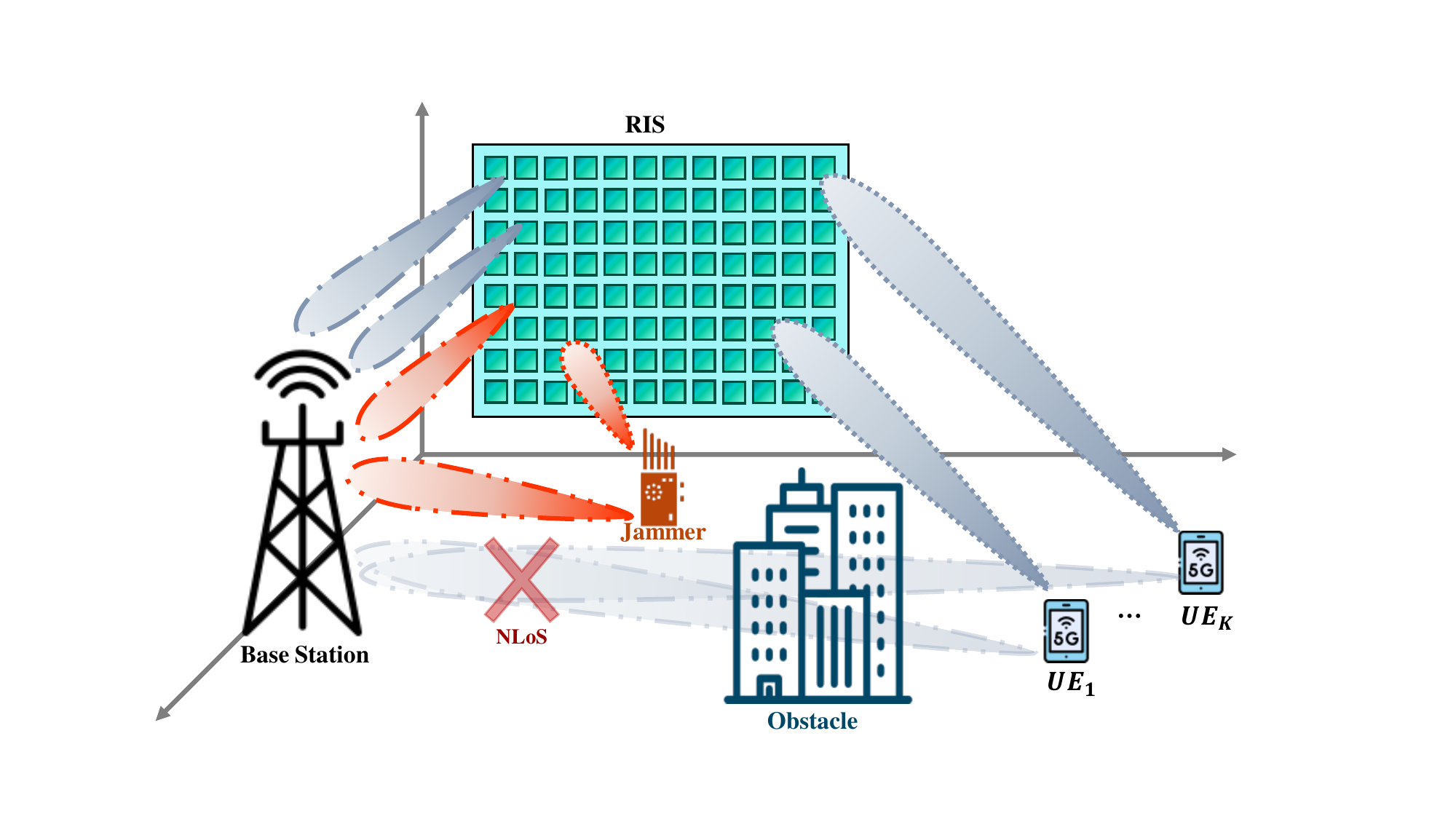}
    \caption{Uplink NOMA communication with $K$ UEs and active RIS under jamming attack.}
    \label{fig:scenario}
    \vspace{-5mm}
 \end{figure}

Consider an uplink NOMA communication including a single-antenna base station or gNodeB and $K$ single-antenna UEs. Due to obstacles in the environment and lack of direct channel links, UEs transmit their signals with the help of RIS as illustrated in Fig. \ref{fig:scenario}. In such a scenario, the transmit signal propagates through two channels of user-RIS and RIS-BS. The RIS in this work is a 2-D rectangle including $N=N_RN_C$ active elements with $N_R$ as the number of elements in each row and $N_C$ as the number of elements in each column. Each element has $d_W = \Delta_h\lambda$ width and $d_H = \Delta_v\lambda$ height assuming $\Delta_h$ and $\Delta_v$ as the horizontal and vertical antenna spacing respectively. $\lambda=\frac{c}{f_c}$ is the wavelength and $c$ denotes the speed of light and $f_c$ represents the central frequency. Therefore, the area of each element can be defined as $A_n = d_W\times d_H, n\in \{1,...,N\}$ and the overall area of RIS is $N\times A_n$. 

\graphicspath{{pics/}}

Considering that the location of the $n$-th antenna element is a matrix of $\boldsymbol{u}_n$ assuming that elements are indexed row by row. $\boldsymbol{u}_n$ can be described as:
\vspace{-0.5mm}
\begin{equation}
    \boldsymbol{u}_n = \begin{bmatrix}
        0\\
        i_R(n)d_W\lambda\\
        i_C(n)d_H\lambda
    \end{bmatrix}.
\end{equation} with $i_R(n)=mod(n-1,N_R)$ and $i_C(n)=\lfloor(n-1)/N_R\rfloor$ as the horizontal and vertical indices of the $\boldsymbol{u}_n$. 
When the antenna array receives the transmit signal, each element of the array located at $\boldsymbol{u}_n$ undergoes the phase shift of $\boldsymbol{\zeta}(\phi^A,\phi^E)^Tu$, where $\phi^A \in (-\pi/2,\pi/2)$ is the azimuth angle and $\phi^E \in (-\pi/2,0)$ is the elevation angle, $\boldsymbol{\zeta}(\phi^A,\phi^E)\in \mathbb{R}^{3\times 1}$ is the wave vector of the incoming signal \cite{bjornson2017massive} described as below:

\begin{equation}
    \boldsymbol{\zeta}(\phi^A,\phi^E)=\frac{2\pi}{\lambda}\begin{bmatrix}\cos{\phi^A}\cos{\phi^E}\\
    \sin{\phi^A}\cos{\phi^E}\\
    \sin{\phi^E}\end{bmatrix}.
\end{equation}

Therefore, the array response $a(\phi^A,\phi^E)\in \mathbb{C}^{N\times 1}$ \cite{bjornson2020rayleigh} can be defined as:
\vspace{-1.5mm}
\begin{equation}\label{eq: arrayRes}
    a(\phi^A,\phi^E) = [e^{j\boldsymbol{\zeta}(\phi^A,\phi^E)^T\boldsymbol{u}_1}, ... , e^{j\zeta(\phi^A,\phi^E)^T\boldsymbol{u}_N}]^T.
\end{equation}

The transmit signal from UE first passes through $RIS-{UE}_k$ channel denoted by $\boldsymbol{G}_k \in \mathbb{C}^{N\times1}$ which is modeled as follows:
\vspace{-2mm}
\begin{equation}
    \boldsymbol{G}_k = \sqrt{Ld_{RIS-{UE}_k}^{-\delta}}e^{-j\frac{2\pi d_{RIS-{UE}_k}}{\lambda}}\times a(\phi^A_k,\phi^E_k),
\end{equation} Here, $L$ is the path gain at the reference distance of $1m$, and $\delta$ is the path loss exponent. $d_{RIS-{UE}_k}$ denotes the distance between RIS and $k$-th UE with azimuth and elevation angles of $\phi_k^A$ and $\phi_k^E$.

After RIS receives the signal from UE it applies active beamforming to the incoming signal using $\boldsymbol{\Theta} \in \mathbb{C}^{N\times1}$ matrix with $\boldsymbol{\Theta}$ defined as $\boldsymbol{\Theta}=diag{(\sqrt{\beta_1}e^{j\theta_1}, ... ,\sqrt{\beta_N}e^{j\theta_N})}$. Here, $\beta_n\in[0,\beta^{max}], n\in\{1,...,N\}$ is the amplitude of the active beamforming with maximum amplitude value of $\beta^{max}$, and $\theta_n\in[0,2\pi],n\in\{1,...,N\}$ is the phase shift of each RIS element.

RIS transmits the UE transmit signal after applying active beamforming to BS using $\boldsymbol{I} \in \mathbb{C}^{N\times1}$ channel which is modelled as:
\begin{equation}\label{eq: RIS-BS}
    \boldsymbol{I}  = \sqrt{Ld_{RIS-BS}^{-\delta}} e^{-j\frac{2\pi d_{RIS-BS}}{\lambda}}\times a(\phi^A_{BS},\phi^E_{BS})
\end{equation} Assuming that RIS is positioned $d_{RIS-BS}$ meter away from the BS and azimuth and elevation angles between RIS and BS are $\phi^A_{BS}$ and $\phi^E_{BS}$. 

Jammer strategically positions itself to optimize the disruption of communication by transmitting jamming signals both directly to the BS and indirectly through RIS. This enhances the effectiveness of the jamming signal without requiring the jammer to consume more energy. The jamming signal passes through the direct channel between the jammer and BS ($h_j$) is defined below assuming that the jammer is  $d_j$ meter away from the BS:
\begin{equation}
h_j = \sqrt{Ld_j^{-\delta}} e^{-j\frac{2\pi d_j}{\lambda}},
\end{equation} 

On the other side, the jamming signal passes through the indirect channel consisting of $RIS-BS$ channel defined in (\ref{eq: RIS-BS}) and $RIS-jammer$ channel ($\boldsymbol{G}_j\in \mathbb{C}^{N\times1}$) described below:
\vspace{-0.3mm}
\begin{equation}
    \boldsymbol{G}_j = \sqrt{Ld_j^{-\delta}} e^{-j\frac{2\pi d_j}{\lambda}}a(\phi^A_j,\phi^E_j),
\end{equation} Where $\phi^A_j$ and $\phi^E_j$ are azimuth and elevation angles between jammer and RIS.

Regarding the SIC decoding order, the BS decodes the signal from ${UE}_1$ to ${UE}_{k-1}$ before the signal from ${UE}_k$.
To meet this requirement, the received signal at the BS from ${UE}_k$ ($1\leq k \leq K$) should be stronger than the signal from ${UE}_{k+1}$ to ${UE}_K$. 
Therefore, the received SJNR of $k$-th UE at the BS under a jamming attack can be described as follows:
\vspace{-0.15mm}
\begin{equation}\label{eq: sinr}
    \gamma_k =
    \frac{P_k|\boldsymbol{I}^T\boldsymbol{\Theta} \boldsymbol{G}_k|^2}{\mathcal{I}+ P_j|h_j + \boldsymbol{I}^T \boldsymbol{\Theta} \boldsymbol{G}_j|^2 + ||\boldsymbol{I}^T\boldsymbol{\Theta}  ||^2\sigma_\upsilon^2+\sigma^2},
\end{equation} Which $k \in \{1,...,K\}$ corresponds to the $k$-th UE in the network. $P_k$ and $P_j$ denote the power of the $k$-th user and jammer respectively.
The interference introduced by the encoded received signals is represented by $\mathcal{I}$ described as $\mathcal{I} = \sum_{k'=k+1}^K P_{k'}|\boldsymbol{I}^T\boldsymbol{\Theta} \boldsymbol{G}_{k'}|^2$.
Given that an active RIS is constructed using electromagnetic components that amplify the signal, it is important to consider the additional noise introduced by these elements in the model \cite{zhang2022active, zhi2022active}. $\boldsymbol{I}^T\boldsymbol{\Theta}\boldsymbol{\upsilon}$ denotes the thermal noise produced by active RIS elements defined as $\boldsymbol{\upsilon}\sim\mathcal{CN}(0,\sigma_\upsilon^2\mathbb{I}^{N\times1})$.
Finally, $z\sim\mathcal{CN}(0,\sigma^2)$ represents additive white Gaussian noise (AWGN) environmental noise at the location of BS. It should be noted that $|.|$ describes the absolute value of a term and $||.||$ denotes the norm of a vector.

\subsection{Network Metrics}
The transmission protocol is based on the transmission diversity along with FBL to meet the requirements of URLLC \cite{polyanskiy2010channel}. UEs are assumed to send their data packets successively through $\mathcal{L}$ number of re-transmissions.
The block error rate (BLER) to decode the received signal at the BS for a blocklength of $n_b$ and $n_d$ information bits per data packet \cite{cao2019joint, abughalwa2022finite} is defined as below:

\begin{equation}
    \epsilon_k \approx Q\left( \sqrt{\frac{n_b}{v(\gamma_k)}} \left(C\left(\gamma_k\right)-\frac{n_d}{n_b} \right) \right),
\end{equation} With $Q(.)$ as the Q-function, $C\left(\gamma_k\right) = \log_2 \left(1+\gamma_k\right)$ as the effective capacity, and  $v(\gamma_k) = \left(1-\frac{1}{(1+\gamma_k)^2}\right)(\log_2e)^2$ as the channel dispersion.

To ensure the URLLC targets are met, main network metrics such as reliability and average delay must be analyzed carefully. Reliability refers to the probability that a typical transmitted packet by ${UE}_k$ is successfully received and decoded at the base station.
Since $\mathcal{L}$ replicas are transmitted for each packet the reliability can be expressed as:
\vspace{-1.5mm}
\begin{equation}
    R_k = \sum_{l=1}^{\mathcal{L}} \binom{l}{\mathcal{L}} (1-\omega_s)^{\mathcal{L}-l}
\end{equation}
\begin{equation}
    R_k =1-[1-\omega_s]^{\mathcal{L}} ,
\end{equation}Where $\omega_s$ is the probability of successfully decoding a single packet replica calculated as:
\vspace{-2mm}
\begin{equation}
    \omega_s = \Pi_{k'=1}^K(1-\epsilon_k')
\end{equation}

The typical time frame in this work is composed of two main components: the payload and the header. The header ($T_h$) contains the information that is essential for the transmission and decoding procedures and the payload ($T_p = \frac{n_b}{B}$) corresponds to the actual data that needs to be transferred where $B$ denotes the bandwidth. The total time taken to transmit both the payload and header is the frame duration $T_f = T_h+T_p$.

The next network metric required to be discussed in URLLC applications is the average delay to transmit a typical packet. A packet spends time waiting in the UE buffer before it is transmitted over the air which is referred to as queuing delay. After transmission, it takes time for the packet to travel from UE to BS in a wireless communication system which is called air latency. The combination of queuing delay and air latency provides the overall packet delay. Consider the transmission diversity model with M/D/1 queuing model in which packets arrive with a Poisson distribution. 
Server utilization for each user ${UE}_k$ is defined as the ratio of the average packet arrival rate of $k$-th user ($\Lambda$) to the average service rate which is expressed as:
\vspace{-2mm}
\begin{equation}
    \rho_k = \frac{\Lambda_k}{\frac{1}{(T_f\times \mathcal{L})}}
\end{equation} Which can be simplified as:
\vspace{-1.5mm}
\begin{equation}
    \rho_k =  \mathcal{L}T_f\Lambda_k
\end{equation}

Therefore, the average transmission delay is as follows \cite{bhat2008introduction}:
\begin{equation}
    \overline{\tau}_k = \mathcal{L}T_f \times \left[\frac{2-\mathcal{L}T_f \Lambda_k}{2(1-\mathcal{L}T_f \Lambda_k)}\right]
\end{equation}

The overall energy efficiency of the system is defined as the total number of bits decoded successfully at the destination to the total consumed energy as expressed below:

\begin{equation} \label{eq: eta}
    \eta = \frac{n_d\sum_{k=1}^K Rel_k}{\sum_{k=1}^K P_k\overline{\tau}_k}
\end{equation}

\subsection{Optimization Problem}\label{sect:opt problem}
The maximization of energy efficiency is carried out by deriving optimal user powers, RIS phase shift and amplitude, blocklength, and maximum number of re-transmissions. Since the RIS receives transmission signals from both users and jammer, it is crucial to determine the optimal settings for $\theta_n$ and $\beta_n$, $n\in\{1,...,N\}$. Therefore, the optimization problem is formulated as follows:

\begin{equation*}
\hspace{-55mm}\underline{\text{\textbf{Energy Efficiency-Max:}}}
\vspace{-2mm}
\end{equation*}
\begin{maxi!}[2]
	{P_k, \theta_n, \beta_n, n_b, \mathcal{L}}{\eta \label{eq:max}}{}{}
	\addConstraint{\overline{\tau}_k}{\leq \overline{\tau}^{thr},\, \, \, \forall k \in \{1,..., K\} \label{eq:consTau}} 
	\addConstraint{Rel_k }{\geq Rel^{thr},  \, \, \, \forall k \in \{1,..., K\}\label{eq:consRel}} 
    \addConstraint{\rho_k<1}{\quad \forall k \in \{1,..., K-1\} \,\label{eq:consRho}}    
	\addConstraint{0 \leq \beta_n}{\leq \beta^{max}},  \, \, \forall n \in \{1,..., N\} \label{eq:consBeta}
    \addConstraint{\theta_n \in \{0,2\pi\}},  \,{\forall n \in \{1,..., N\} \label{eq:consTheta}}
    \addConstraint{\mathcal{L}}{\in \{1,...,\mathcal{L}_{max}\} \label{eq:consLmax}}
    \addConstraint{0< P_{k}}{\leq P_{k+1} \, \, \, \forall k \in \{1,..., K-1\} \label{eq:consP}}
    \addConstraint{P_{k}}{\leq P_{max} \, \, \, \forall k \in \{1,..., K\} \label{eq:consPmax}}
	\addConstraint{n_b}{\in \mathbb{N} .\label{eq:consNb}}
\end{maxi!}With constraint (\ref{eq:consTau}) denoting the latency requirement for URLLC application, constraint (\ref{eq:consRel}) ensuring the URLLC reliability is met, constraint (\ref{eq:consRho}) is related to the stability of service utilization, constraint (\ref{eq:consBeta}) denotes the active RIS capability to amplify, reflect, or absorb the incoming signal, constraints (\ref{eq:consTheta}) and (\ref{eq:consLmax}) define the lower and upper bounds of RIS phase shift values and the maximum number of retransmissions respectively, constraint (\ref{eq:consP}) reflects the SIC decoding order, constraint (\ref{eq:consPmax}) ensures that the assigned power to the users do not exceed the maximum, and constraint (\ref{eq:consNb}) is related to the positive integer nature of the blocklength.

\section{Numerical Results} \label{Sec:Numerical_result}

In this section, the performance of the RIS-based NOMA system model demonstrated in Fig.\ref{fig:scenario} is investigated using MATLAB. Two users ($K=2$) send packets to the BS through a RIS\footnote{The two-node NOMA communication is an elementary block of NOMA, addressed in 3GPP \cite{3GPPTR36.859}.}. The RIS is positioned at (0,0) in a Cartesian coordinate system with a different number of elements from $N=4$ to $N=900$ \cite{ge2022ris}. The central frequency is $f_c=28\ GHz$ and the bandwidth is $B=180\ KHz$. Environmental noise and dynamic noise are $\sigma_\upsilon^2=\sigma^2=-100\ dBm$ \cite{zhang2022active}. Users have similar arrival rates ($\Lambda_1 = \Lambda_2 = \Lambda$) which results in the same average delays ($\overline{\tau}_1 = \overline{\tau}_2 = \overline{\tau}$). Table \ref{table:parameters} provides the summary of simulation parameters inspired by \cite{ haghshenas2023new}.

\begin{table}[h!] 
\centering
\renewcommand{\arraystretch}{1.45}
\caption{Simulation configuration }
\begin{tabular}{||m{4.7em}| m{1.5cm} || m{4.7em}| m{1.90cm} ||} 
 \hline
 Parameter & Value & Parameter & Value\\ [0.5ex] 
 \hline\hline
   $\Delta_h, \Delta_v$ & $0.25$ & $N_R, N_C$ & $0.25$\\
   \hline
    $\lambda$ & $0.0107\ m$ & $\Lambda$ & $500$ \\
   \hline
   $\delta$ & $2$ & $L$ & $30\ dB$ \\
   \hline
    $\phi^A_{BS}$ & $\pi/6$ & $\phi^E_{BS}$ & 0\\
   \hline
   $d_{RIS-BS}$ & $4\ m$ & $d_{RIS-UE_1}$ & $20\ m$\\
   \hline
   $\phi^A_k$ & $\pi/2$ & $\phi^E_k$ & $2\pi$\\
   \hline
    $d_{RIS-UE_2}$ & $25\ m$ & $T_h$ & $30\ \mu s$ \\
   \hline
    $n_d$ & $32\ B$ & $\mathcal{L}$ & $10$\\
   \hline
    $P_j$ & $5\ mW$ & $d_j$ & $30\ m$ \\
    \hline
    $\phi^A_j$ & $\pi/4$ & $\phi^E_j$ & $\pi/2$\\
   [1.5ex] 
 \hline
\end{tabular}
\label{table:parameters}
\vspace{-3mm}
\end{table}

The optimization problem is solved using a genetic algorithm.
As the solution space is large due to the high number of parameters, a large population size is required to increase the probability of covering a wide search space which enhances the probability of finding the optimal solution \footnote{The reader can refer to \cite{asemian2025active} which addresses the limitations of genetic algorithm using surrogate optimization and deep learning}. Meanwhile, the algorithm requires a sufficient number of iterations to converge toward the optimal solution while considering the computational efficiency.
Therefore, each population has a size of $2000$ and goes through the algorithm with $200$ maximum number of generations. These values are selected empirically during the preliminary experiments.
The algorithm minimizes the objective function of $f(x) = \frac{1}{\eta}$. The constraint and function tolerance values are set to $10^{-30}$ to guarantee that constraints are satisfied. Maximum values of $\beta_{max} = 100\ mW$ and $P_{max}=100\ mW$ are also considered to avoid excessive power consumption. The maximum number of allowed re-transmissions is set to $\mathcal{L}_{max}=10$ according to \cite{3gpp.38.321}.
The latency ($\overline{\tau}^{thr}=1\ ms$) and reliability ($Rel^{thr}=0.99999$) threshold values for URLLC requirements are based on the 3GPP standard \cite{3gpp.38.913}.

\begin{figure}[htbp]
    \centering
    \vspace{-5mm}
    \includegraphics[trim={9cm 3cm 9cm 4cm},clip , width=0.85\linewidth, height = 5cm]{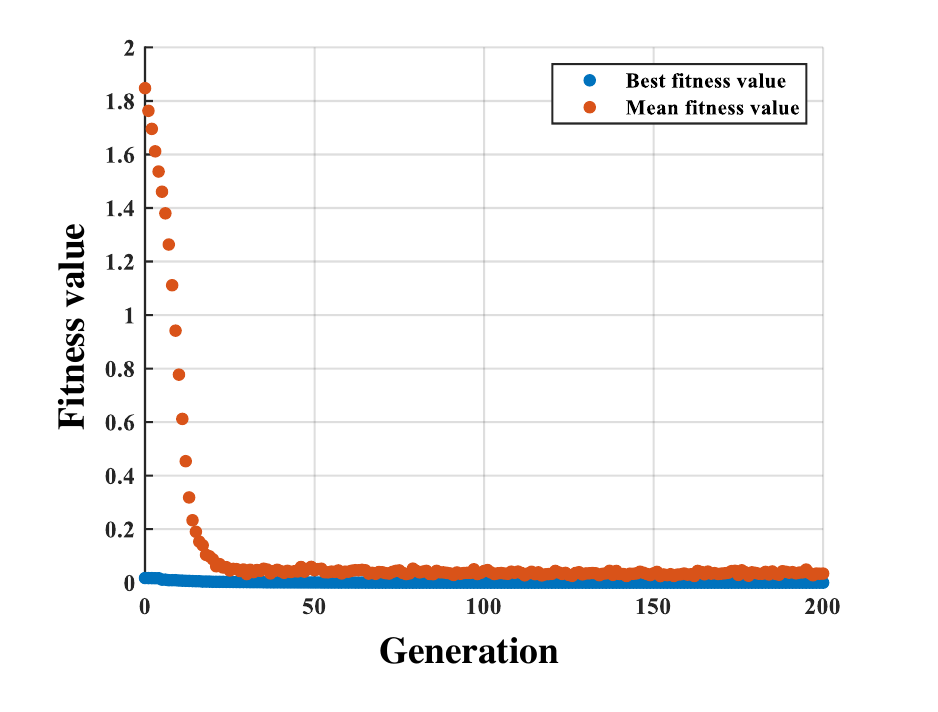}
    \caption{Genetic algorithm convergence in the optimization.}
    \label{fig:ga}
 \end{figure}

The convergence of the genetic algorithm over $200$ generations is illustrated in Fig. \ref{fig:ga}. The figure shows the evolution of solutions toward an optimal point by minimizing $1/\eta$.

\begin{figure}[htbp]
    \centering
    \begin{subfigure}{0.4\textwidth}
        \includegraphics[trim={9.7cm 3cm 10cm 5.08cm},clip , width=0.9\linewidth]{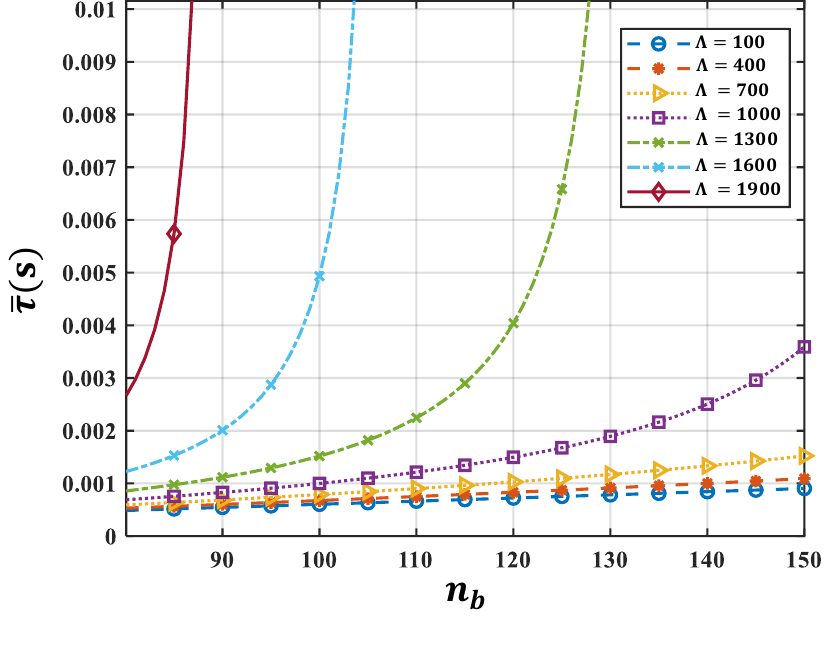}
        \caption{} 
        \label{fig:tau_nb_Lambda}
    \end{subfigure}
    \hfill
    \begin{subfigure}{0.4\textwidth}
         \centering
         \includegraphics[trim={10cm 3cm 10cm 4.5cm},clip , width=0.9\linewidth]{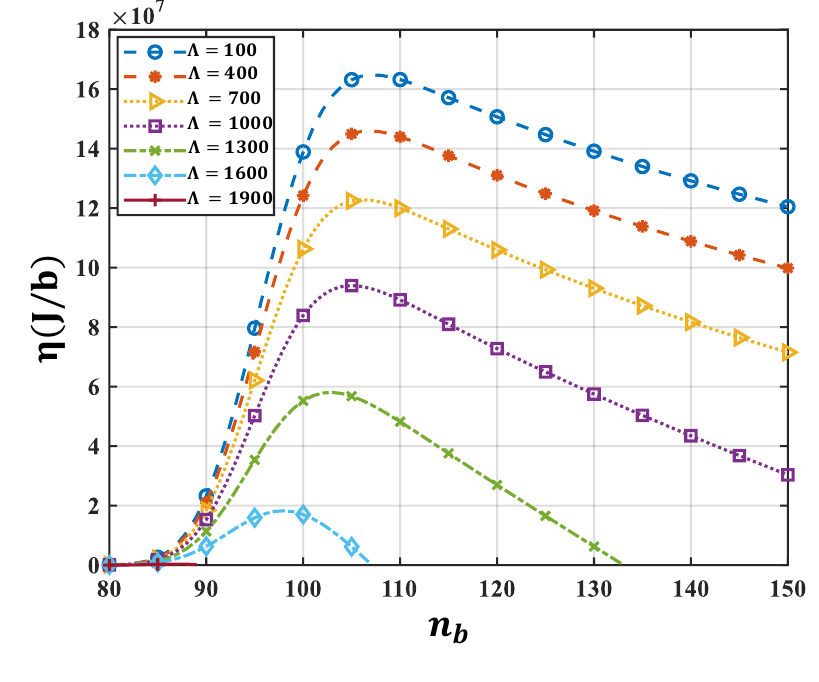}
         \caption{}
        \label{fig:eta_nb_Lambda}
    \end{subfigure}
    \caption{Network metrics under different arrival rates vs different blocklenghths.}
    \label{fig: nb_Lambda}
\end{figure}

\graphicspath{{pics/}}
\vspace{-2mm}
Fig. \ref{fig:tau_nb_Lambda} demonstrates the average delay when blocklength increases for different arrival rate values.
For lower $\Lambda$ values $\overline{\tau}$ and $n_b$ have a linear behaviour as the significant factor of latency is over the air transmission related to the frame length. As $\Lambda$ increases, the role of the waiting time in the buffer of the transmitter gains importance and becomes a determinant element in the latency. Thus, a non-linear behaviour is observed for $\overline{\tau}$ and $n_b$ as $\Lambda$ increases.

Fig. \ref{fig:eta_nb_Lambda} shows the energy efficiency versus blocklength for different arrival rate settings. For lower blocklength values ($n_b$ lower than 90), reliability has a low value. Thus, the number of error-free decoded bits are lower which results in lower energy efficiency. As $n_b$ increases, energy efficiency experiences a growing behaviour to reach its maximum value (for instance for $\Lambda=100$ and $n_b=108$ energy efficiency reaches $\eta=16\times10^7$). At this point, reliability has achieved its highest values and increasing the blocklength has a counterproductive effect on $\eta$. This is due to the fact that increasing the blocklength results in higher frame length and therefore the consumed energy for a constant number of bits is increased.

The importance of the parameter optimization is obvious in Fig. \ref{fig: nb_Lambda}. With an arrival rate of $\Lambda=100$ and $n_b=108$ the average delay and the energy efficiency is $31.68\%$ lower and higher than $\Lambda=1300$ and $n_b=108$ respectively (the average delay is $6.51\times10^{-4}$ for $\Lambda=100$ and $0.0021$ for $\Lambda=1300$).
\vspace{-0.5mm}
\graphicspath{{pics/}}
\vspace{-1mm}

\begin{figure}[htbp]
    \centering
    \includegraphics[trim={10cm 3cm 10cm 4.85cm},clip , width=0.9\linewidth, height=6cm]{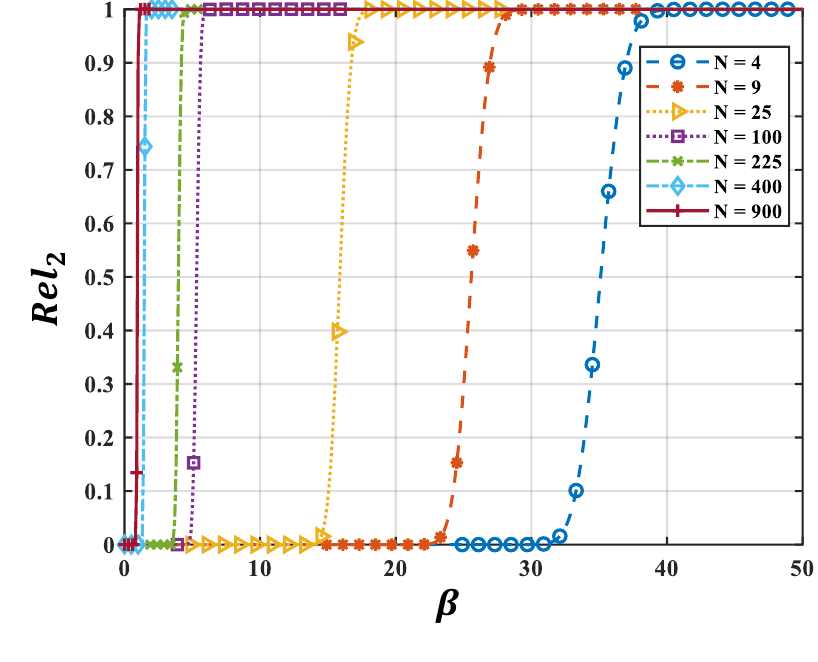}
    \vspace{-3mm}
    \caption{Reliability corresponding to the transmission of the second user as a function of RIS elements amplitude for different numbers of RIS elements.}
    \label{fig:rel vs beta}
 \end{figure}

\begin{figure}[htbp]
    \centering
    \includegraphics[trim={10cm 3cm 9.5cm 4.7cm},clip , width=0.7\linewidth, height = 5.5cm]{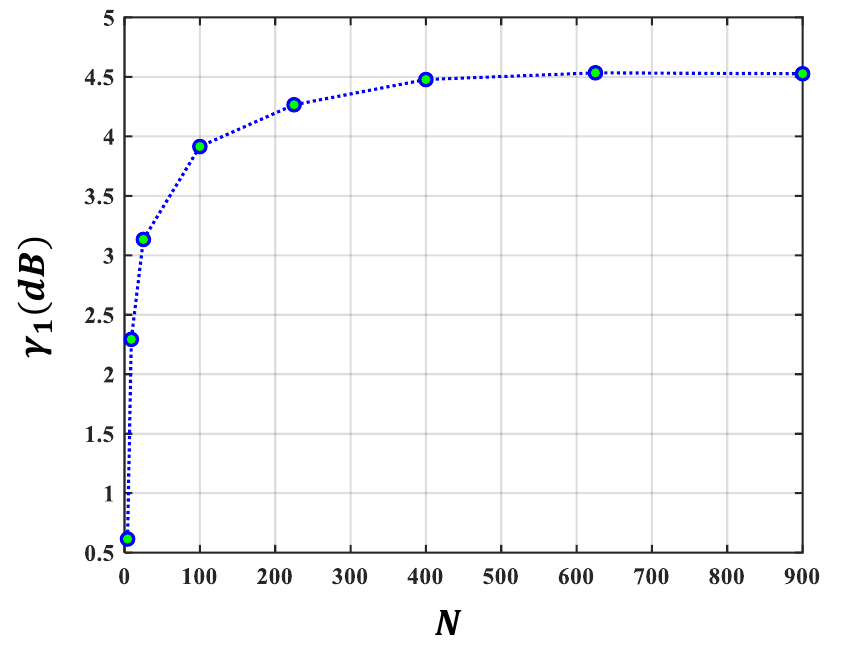}
    \caption{The impact of the number of RIS elements on the SJNR for the first user.}
    \label{fig:gamma vs N}
    \vspace{-5mm}
 \end{figure}

Fig. \ref{fig:rel vs beta} illustrates the impact of the amplitude of the active RIS elements on the reliability of transmission for the second user. Each curve is related to setups with a different number of RIS elements. Based on the figure, as the amplitude of the RIS increases, the reliability is enhanced. Additionally, for $N=4$ the system achieves the desired reliability at $\beta=43.7$ while for $N=400$ the URLLC reliability is achieved at $\beta=2.1$. Thus, the higher number of RIS elements to $N=400$ achieves better reliability at significantly lower amplitude which reduces the power consumption of the RIS.
Therefore, on an optimal setup, the system reaches the required URLLC reliability with amplitudes below $50$, thus, high values of amplitude are not necessary to enhance the performance of the network.
It should be noted that increasing the number of elements from $N=400$ to $N=900$ results in only marginal improvements in reliability, indicating that the system has achieved a plateau in performance. This is furthermore discussed by analyzing the SJNR in Fig. \ref{fig:gamma vs N}.
\vspace{-1mm}
\graphicspath{{pics/}}
\vspace{-2mm}
Considering (\ref{eq: eta}) it can be concluded that SJNR is a critical metric that directly impacts the energy efficiency of the network. Higher SJNR values mean that the system is able to decode more bits without requiring re-transmission which leads to lower BLER and thus higher reliability. In Fig. \ref{fig:gamma vs N} the relationship between the SJNR and the number of RIS elements for the first user is demonstrated. From $N=4$ to $N=400$ the SJNR increases rapidly from $0.61$ to $4.47$. As the number of elements continues to increase to $N=400$, the rate of the growth of SJNR becomes a plateau. This is mainly due to the interference effect of active RIS elements ($\sigma_\upsilon^2$ in (\ref{eq: sinr})). As the number of elements increases the thermal noise produced by the active elements increases and it no longer improves the SJNR and therefore energy efficiency.
 \vspace{-1mm}
 \graphicspath{{pics/}}
 
\section{Conclusions} \label{Sec:Conclusions}
This study has addressed the challenge of jamming attacks in NOMA-based 5G networks that operate under FBL conditions. Our proposal combines RIS active elements to improve energy efficiency while simultaneously satisfying the reliability and latency of URLLC applications. Through results from several simulations, we show how the corporation of optimal parameters mitigates the effect of jammer and improves network metrics such as reliability, latency, and energy efficiency. we also explored the impact of the number of RIS elements and it is concluded that increasing the number of RIS elements improves SJNR which directly impacts energy efficiency. However, there is an optimal point beyond which they plateau due to the increased thermal noise from the active elements. The genetic algorithm applied to optimize the system is efficient in navigating the complex solution space. 


\section*{Acknowledgment}

This work was supported in part by funding from the Innovation for Defence Excellence and Security (IDEaS) program from the Department of National Defence (DND) and the Natural Science and Engineering Research Council (NSERC) CREATE TRAVERSAL Program

\bibliographystyle{IEEEtran}

\end{document}